\documentstyle[amssymb,cite]{elsart}
\input epsf
\epsfclipon
\begin{document}
\begin{frontmatter}
\title{ Higher order clustering coefficients in Barab\'{a}si-Albert networks }
\author{Agata Fronczak},
\author{Janusz A. Ho\l yst \thanksref{emjh}},
\author{Maciej Jedynak},
\author{Julian Sienkiewicz}
\address{Faculty   of   Physics,   Warsaw University
of Technology, Koszykowa 75, \\ \hspace*{0.15em} PL-00--662
Warsaw, Poland }
\thanks[emjh]{e-mail: jholyst@if.pw.edu.pl}
\date{\today}
\begin{abstract}
Higher order clustering coefficients $C(x)$ are introduced for
random networks. The coefficients express probabilities that the
shortest distance between any two nearest neighbours of a certain
vertex $i$ equals $x$, when one neglects all paths crossing the
node $i$. Using $C(x)$ we found that in the Barab\'{a}si-Albert
(BA) model the average shortest path length in a node's
neighbourhood is smaller than the equivalent quantity of the whole
network and the remainder depends only on the network parameter
$m$. Our results show that small values of the standard clustering
coefficient in large BA networks are due to random character of
the nearest neighbourhood of vertices in such networks.
\end{abstract}
\end{frontmatter}
\newpage
%
\section{Introduction}

During the last few years studies of random, evolving networks (
such as the Internet, WWW, social networks, metabolic networks,
food webs etc - for review see \cite{przegl1, przegl2}) have
become a very popular research domain among physicists. A lot of
efforts were put into investigation of such systems in order to
recognize their structure and to analyse emerging complex
dynamics. We learned that networks are far from being random as
Erd\"{o}s and R\'{e}nyi assumed in random graph theory
\cite{RG1,RG2}, but surely they are not so ordered as crystals.
Despite network diversity, most of real weblike systems share
three prominent features \cite{przegl1,przegl2}:
\begin{itemize}
\item
The average shortest path length $l$ is small. In order to connect
two nodes in a network typically only a few edges need to be
passed.
\item
The average clustering coefficient $C$ is large. Two nodes having
a common neighbour are also likely to be neighbours.
\item
The probability that a randomly selected node has exactly $k$
nearest neighbours follows a power law (scale-free) distribution
$P(k)\sim k^{-\gamma}$ with $2<\gamma<3$ in most of real systems.
\end{itemize}

A considerable number of network models has been studied in order
to capture the above characteristics. Most of these are based on
two ingredients originally introduced by Barab\'{a}si and Albert
\cite{ba1,ba2}: continuous network growth and preferential
attachment. In Barab\'{a}si-Albert (BA) model, network starts to
grow from an initial cluster of $m$ fully connected sites. Each
new node that is added to the network creates $m$ links that
connect it to previously added nodes. The preferential attachment
means that the probability of a new link to end up in a vertex $i$
is proportional to connectivity $k_{i}$ of this vertex. The
validity of the preferential attachment was confirmed within real
networks analyses \cite{bareal, new1, new2}. The BA algorithm
generates networks with the desirable scale-free distribution
$P(k)\sim k^{-3}$ and small values of the average shortest path.
One can also observe a phase transition for spins located at BA
network vertices with a critical temperature increasing as a
logarithm of the system size \cite{ahs, bianconi, isingRG,
isingIT}. The only striking discrepancy between the BA model and
real networks is that the value of the clustering coefficient
predicted by the theoretical model decays very fast with network
size and for large systems is typically several orders of
magnitude lower than found empirically.

In this paper we extend the standard definition of the clustering
coefficient by introducing higher order clustering coefficients
that describe interrelations between vertices belonging to the
nearest neighbourhood of a certain vertex in complex network.
Global characteristics like the standard clustering coefficient
and the average shortest path do not provide a useful insight into
complex network structure and dynamics. We hope that the higher
order clustering coefficient analyses in real systems
\cite{afreal} may give some guidelines how to model clustering
mechanisms. Here, we study higher order clustering coefficients in
BA model. Our results provide a vivid evidence that the BA
networks are blind to clustering mechanisms.
%
%
\section{Model description}

The standard clustering coefficient $C$ is one of global
parameters used to characterise the topology of complex networks.
For the first time it was introduced by Watts and Strogatz
\cite{watts} to characterise local transitivity in social
networks. Clustering coefficient gives the probability that two
nearest neighbours of the same node are also mutual neighbours.
Let us focus on a selected node $i$ in a network, having $k_{i}$
edges which connect it to $k_{i}$ other nodes. The value of the
clustering coefficient of the node $i$ is given by the ratio
between the number of edges $E_{i}$ that actually exist between
these $k_{i}$ nodes and the total number $k_{i}(k_{i}-1)/2$ of
such edges that could exist in the neighbourhood of $i$:
\begin{equation}\label{eq1}
C_{i}=\frac{2E_{i}}{k_{i}(k_{i}-1)}.
\end{equation}
The clustering coefficient of the whole network is the average of
all individual $C_{i}$'s

We define a {\it clustering coefficient of order $x$} for a node
$i$ as the probability that there is a distance of length $x$
between two neighbours of a node $i$. Putting the number of such
$x$-distances equal to $E_{i}(x)$, the higher order clustering
coefficients follow:
\begin{equation} \label{eq2}
C_{i}(x)=\frac{2E_{i}(x)}{k_{i}(k_{i}-1)}.
\end{equation}
$C(x)$ is the mean value of $C_{i}(x)$ over the whole network.
Note that $C(x)$ reduces to the standard clustering coefficient
for $x=1$ and $\sum_{x} C(x)=1$ for the BA networks with $m\geq
2$.

\begin{figure}[h]
\epsfxsize=15cm \epsfbox{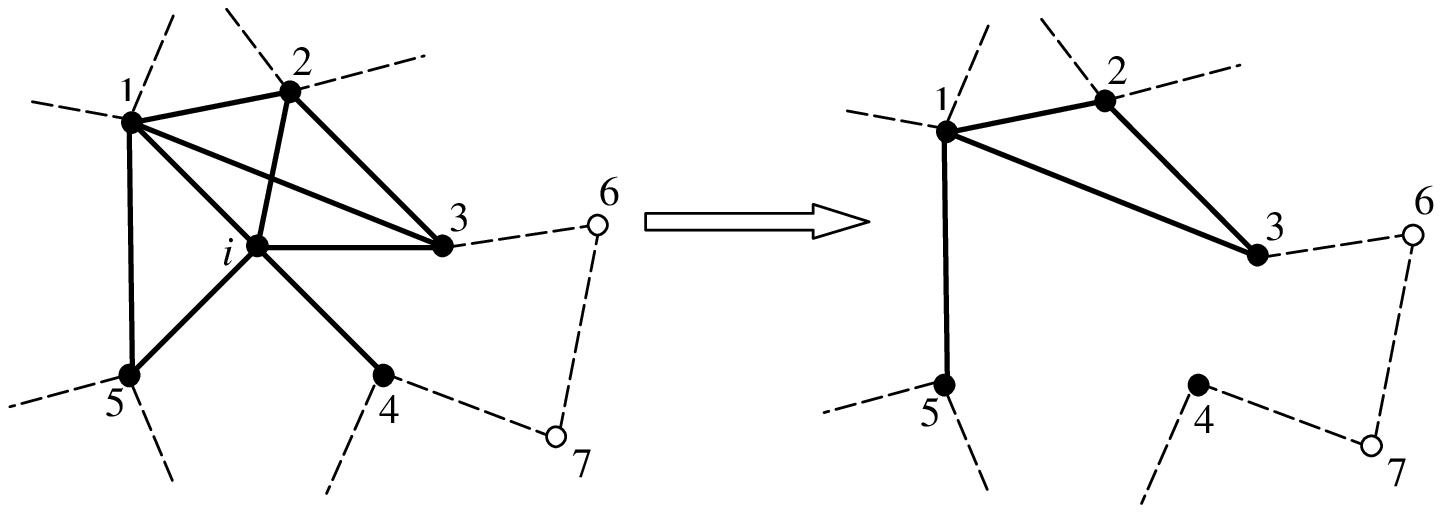} \vskip 1cm \caption{{\it Stage
1:} Network in the vicinity of a node $i$. {\it Stage 2:} After
removing the node $i$ and its adjacent links.} \label{fig1}
\end{figure}

The above definition becomes comprehensible after examining
Fig.\ref{fig1}. Let us assume a node $i$ with $k_{i}=5$. The node
determines its nearest neighbourhood that in this case consists of
vertices $\{1,2,3,4,5\}$. The aim is to find higher order
clustering coefficients $C_{i}(x)$ of the node $i$.
%
%
The table below gives the shortest distances between vertices
adjacent to the node $i$. The distances were taken from the
Fig.\ref{fig1} ({\it Stage $2$}):
%
\[\bordermatrix{&1 &2 &3 &4 &5 &\cr 1 &- & 1 &1 &4 &1 \cr 2 &1 &- &1 &4 &2 \cr 3 &1 &1 &- &3 &2 \cr 4 &4 &4 &3 &- &5 \cr
5 &1 &2 &2 &5 &- \cr}.\]
Summing over all pairs of vertices one obtains:
%
\[\bordermatrix{ &\cr x &1 &2 &3 &4 &5 \cr E_{i}(x) &4 &2 &1 &2 &1 \cr C_{i}(x) &0.4 &0.2 &0.1 &0.2 &0.1
\cr}.\]
%
%
%
\section{Numerical results}

\begin{figure}[h]
\epsfxsize=12cm \epsfbox{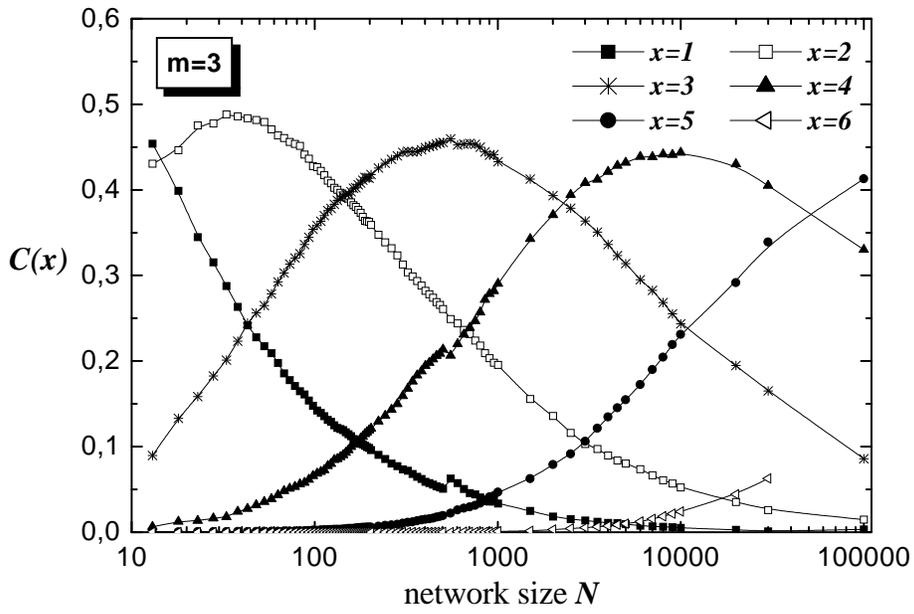} \vskip 2cm \epsfxsize=12cm
\epsfbox{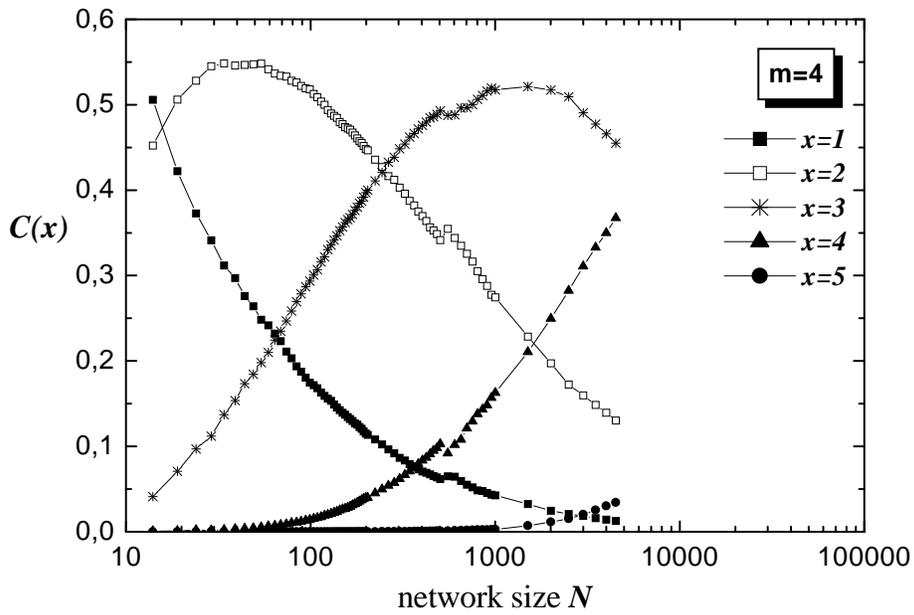} \caption{Higher order clustering coefficients
versus network size $N$ for $m=3$ and $4$.} \label{fig2}
\end{figure}

Fig.\ref{fig2} shows the higher order clustering coefficients
dependence on the network size $N$. Nodes determined as belonging
to the nearest neighbourhood of any vertex in BA network are more
likely to be second ($x=2$), third ($x=3$) and further neighbours
when $N$ increases.
%
%

\begin{figure} [h]
\epsfxsize=12cm \epsfbox{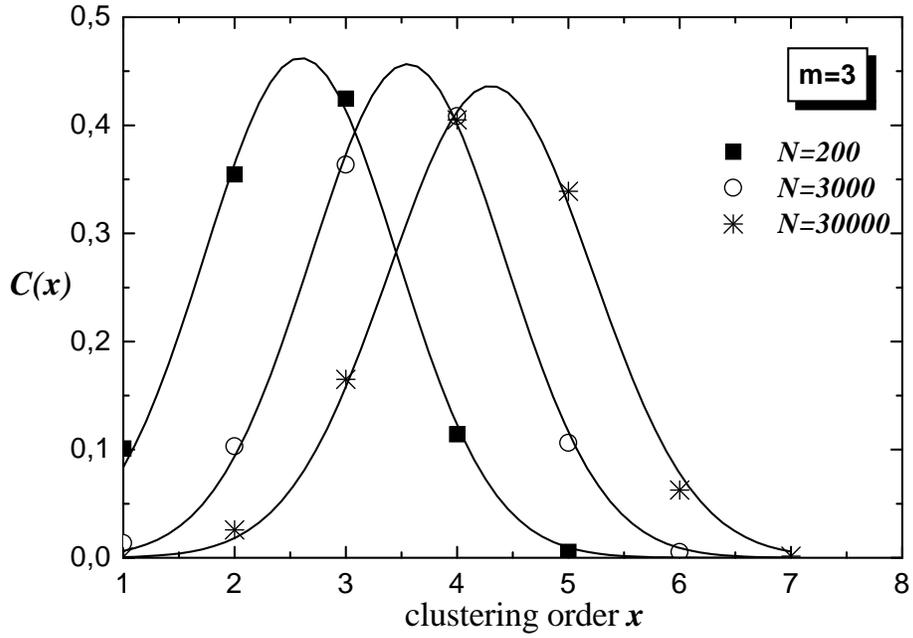} \vskip 2cm \epsfxsize=12cm
\epsfbox{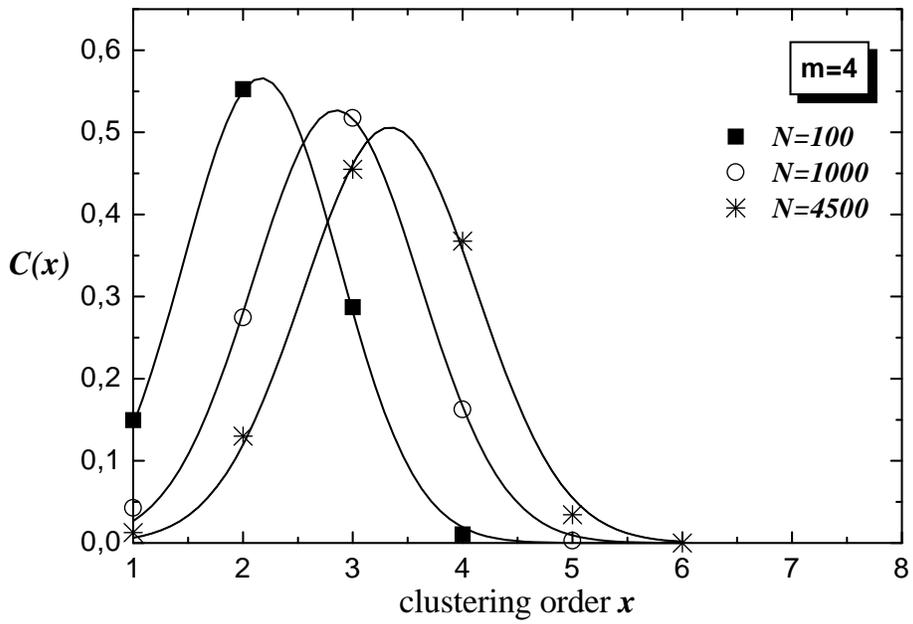} \caption{Higher order clustering coefficient
distributions in BA networks of given sizes $N$ for $m=3$ and
$4$.} \label{fig3}
\end{figure}

We investigated distributions of the clustering coefficients
$C(x)$ at a given system size $N$ in BA networks with $m=3$ and
$4$. We found that regardless of $N$ the distributions of $C(x)$
(when extending $x$ to real values) fairly good fit the normalized
Gaussian curve (Fig.\ref{fig3}). Moreover, we observed that the
standard deviation of these distributions depends only on the
parameter $m$ of the network. The gaussian-like patterns of $C(x)$
in BA model express {\it random} character of clustering
relationships. In fact, it is known that the distribution of
distances between randomly chosen sites in BA network is Gaussian.
It follows that interrelations in the nearest neighbourhood of any
vertex in BA model are similar to interrelations between randomly
chosen vertices in the BA network. Distinctly, there are no
mechanisms responsible for clustering in Barab\'{a}si-Albert
algorithm.
%
%

\begin{figure}[h]
\epsfxsize=12cm \epsfbox{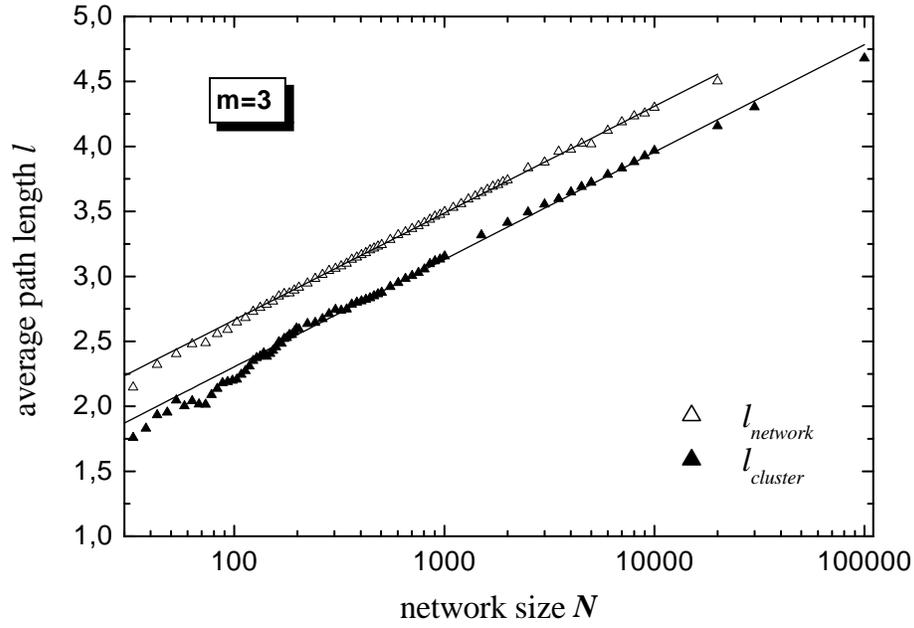} \vskip 2cm \epsfxsize=12cm
\epsfbox{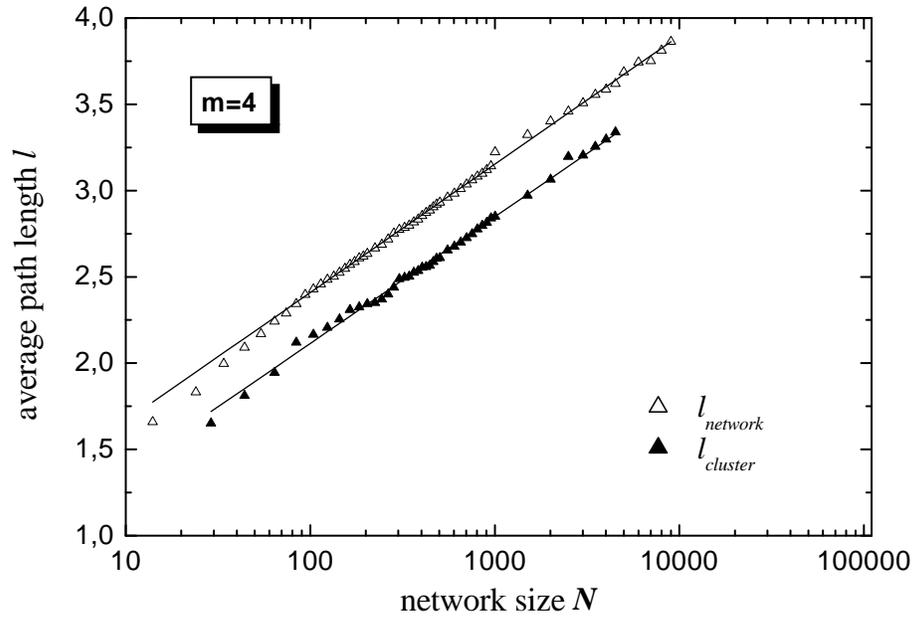} \caption{Characteristic path lengths
$l_{cluster}$ and $l_{network}$ versus network size $N$ for $m=3$
and $4$. For $m=3$: $l_{network}$ is fitted to $0.82 \ln(N)+1.02$
and  $l_{cluster}$ is fitted to $0.83 \ln(N)+0.65$. For $m=4$:
$l_{network}$ is fitted to $0.74 \ln(N)+0.92$ and $l_{cluster}$ is
fitted to $0.73 \ln(N)+0.65$.} \label{fig4}
\end{figure}

The above-described observation can be verified by a functional
dependence of centers $x_{c}$ of the distribution $C(x)$ on the
network size $N$. We realised that the value $x_{c}$
(Fig.\ref{fig3}) expresses both: the order $x$ when $C(x)$ obtains
maximum and the average shortest path length $l_{cluster}$ between
vertices belonging to the neighbourhood of any vertex within a
network. Fig.\ref{fig4} shows that in BA networks the average
shortest path $l_{cluster}$ scales with the system size $N$ in the
same way as the average shortest path length for the whole network
$l_{network}$ \cite{przegl1}, i.e. in the first approximation as:
\begin{equation}\label{l1}
l_{cluster} \sim \ln(N).
\end{equation}
%
%

We found that the mean distance between two nearest neighbours of
the same node $l_{cluster}$ equals to the mean distance between
any two nodes in the network $l_{network}$ minus a constant $A(m)$
\begin{equation}\label{l2}
l_{cluster}=l_{network}-A(m),
\end{equation}
where $A(m)$ is approximately independent on the network size $N$
and equals $0.37$ and $0.27$ for $m=3$ and $m=4$ respectively.
%
%
\section{Conclusions}

In summary, we quantified the structural properties of BA networks
by the higher order clustering coefficients $C(x)$ defined as
probabilities that the shortest distance between any two nearest
neighbours of a certain vertex $i$ equals $x$, when neglecting all
paths crossing the node $i$. We estimated that the average
shortest path length in the node's neighbourhood is smaller than
the equivalent whole network quantity and the remainder depends
only on the network parameter $m$. Our results show in a vivid way
that the absence of the clustering phenomenon in BA networks is
due to the random character of the nearest neighbourhood in these
networks.

Recently some alternative algorithms have been suggested to
account for the high clustering found in real weblike systems.
Holm and Kim \cite{Holme} have extended the standard BA model
adding the trial formation rule. Networks built according to their
guidelines exhibit both the high clustering and the scale-free
nature. Barabasi et al. \cite{bhier1,bhier2} and independently
Dorogovtsev et al. \cite{dhier} have found that scale-free random
networks can be modelled in a deterministic manner by so-called
pseudofractal scale-free networks. They argued that the high
clustering in real networks might result from their hierarchical
topology. It would be interesting to analyse the higher order
clustering coefficients in these networks and compare it with real
data \cite{afreal}.
\newpage
%
%

%
%
\newpage
\end{document}